\documentclass[twocolumn,floatfix,eqsecnum,aps,pre]{revtex4-2}
\usepackage{graphicx}
\usepackage{amsmath}
\usepackage{amssymb}
\usepackage{bm}
\usepackage{hyperref}

\usepackage{color}

\begin{document}

\title{Entropy and singular-value moments\\
of products of truncated random unitary matrices}
\author{C. W. J. Beenakker}
\affiliation{Instituut-Lorentz, Universiteit Leiden, P.O. Box 9506, 2300 RA Leiden, The Netherlands}

\date{January 2025}

\begin{abstract}
Products of truncated unitary matrices, independently and uniformly drawn from the unitary group, can be used to study universal aspects of monitored quantum circuits. The von Neumann entropy of the corresponding density matrix decreases with increasing length $L$ of the product chain, in a way that depends on the matrix dimension  $N$ and the truncation depth $\delta N$. Here we study that dependence in the double-scaling limit $L,N\rightarrow\infty$, at fixed ratio $\tau=L\delta N/N$. The entropy reduction crosses over from a linear to a logarithmic dependence on $\tau$ when this parameter crosses unity. The central technical result is an expression for the singular-value moments of the matrix product in terms of the Erlang function from queueing theory.
\end{abstract}
\maketitle

\section{Introduction}

Matrix contraction processes \cite{Wil18}, in which sub-unitary matrices are multiplied to reduce the rank of the matrix product, were introduced by von Neumann \cite{Neu49,Hal62,Gin18} and have become an active topic in random-matrix theory \cite{Cri93,Ake14,Ake15,Ibs14,Adh16,Kie16,For18,Adh20,Gia23}. A recent application in the context of quantum information processing is to monitored quantum circuits \cite{Bul24,Ger24,DeL24}: The unitary evolution in an $N$-dimensional Hilbert space is alternated by projective measurements on an $N-\delta N$ dimensional subspace. An initially mixed density matrix $\rho$, of full rank, approaches a rank-one pure state after a sufficiently large number $L$ of measurements \cite{Li18,Cha19,Ski19,Szy19,Cho20,Gul20}, as quantified by the decreasing von Neumann entropy $-\operatorname{Tr}\rho\ln\rho$.

When $L\gg 1$ the purification happens even for weak non-unitarity, with $\delta N$ of order unity while $N\gg 1$. Here we study this double-scaling limit, $L,N\rightarrow\infty$ at fixed ratio $\tau=L\delta N/N$, which allows for a universal single-parameter characterization if the unitary matrices are drawn independently and uniformly from the unitary group ${\rm U}(N)$. The universality ensures that our results are consistent with those obtained for different models of weak measurements \cite{Bul24,Ger24,DeL24}. The merit of our alternative approach is its simplicity, we only rely on elementary methods of Gaussian averages.

The outline of this paper is as follows. In the next section we formulate the random contraction process, with its large-$N$ Gaussian limit. In that limit the singular-value moments satisfy a closed set of recursion relations, that we present and solve in Sec.\ \ref{sec_recursion}, for finite $L$, before taking the double-scaling limit $N,L\rightarrow\infty$ in Sec. \ref{sec_scaling}. The probability density of the singular values and the corresponding von Neumann entropy are calculated in Secs.\ \ref{sec_density} and \ref{sec_entropy}. We conclude in Sec.\ \ref{sec_conclude}.

\section{Random contractions}

\subsection{Formulation as a product of truncated Haar-distributed matrices}

We consider a set of $L$ unitary matrices $U_1,U_2,\ldots U_L$, each of size $N\times N$, drawn independently and uniformly according to the Haar measure on the unitary group $\mathrm{U}(N)$. Our analysis will also apply to the case of real Haar-distributed matrices from the orthogonal group $\mathrm{O}(N)$. (In the latter case, the Hermitian conjugate $U_n^\dagger$ reduces to the transpose.)

Each matrix $U_n$ is truncated by zeroing out the first $\delta N$ rows and columns \cite{note1},
\begin{equation}
\tilde{U}_n= {\cal P}U_n {\cal P},\;\;{\cal P}_{ij}=\begin{cases}
1&\text{if}\;\;i=j>\delta N,\\
0&\text{otherwise}.
\end{cases}
\end{equation}

We construct a random contraction from the matrix product
\begin{equation}
C_L=\tilde{U}_L\tilde{U}_{L-1}\cdots\tilde{U}_2\tilde{U}_1.\label{CLdef}
\end{equation}
The singular values squared $\sigma^2$ (the eigenvalues of $C_L^{\vphantom\dagger}C_L^\dagger$), all lie in the interval $[0,1]$. We seek their statistical moments,
\begin{equation}
\mathbb{E}[\sigma^{2p}]=\frac{1}{N}\,\mathbb{E}\bigl[\operatorname{Tr}(C_L^{\vphantom\dagger}C_L^\dagger)^p\bigr].
\end{equation}

\subsection{Formulation as a random projector product}

An equivalent formulation is in terms of random projections. Collect the first $\delta N$ columns of $U_n$ into the orthonormal set of vectors $u_1^{(n)},u_2^{(n)},\ldots u_{\delta N}^{(n)}$ and define the projector
\begin{equation}
P_n=\openone-\sum_{i=1}^{\delta N}|u_i^{(n)}\rangle\langle u_i^{(n)}|,\;\;P_n^2=P_n.
\end{equation}
The projector product
\begin{equation}
Q_L=P_L P_{L-1}\cdots P_2 P_1\label{NLdef}
\end{equation}
has the same singular-value statistics as $C_{L-1}$, a correspondence noted recently \cite{Bul24}.

To see the equivalence of the two types of random contractions, note that, by construction, $P_n= U_n^{\vphantom{\dagger}} {\cal P} U_n^\dagger$, hence Eq.\ \eqref{NLdef} can equivalently be written as $Q_L=U_L^{\vphantom{\dagger}}Q'_LU_1^\dagger$ with
\begin{subequations}
\begin{align}
Q'_L={}& {\cal P} U_L^\dagger U_{L-1}^{\vphantom{\dagger}} {\cal P} U_{L-1}^\dagger\cdots U_2^{\vphantom{\dagger}} {\cal P} U_2^\dagger U_1^{\vphantom{\dagger}} {\cal P} \nonumber\\
={}&\tilde{V}_{L-1}\tilde{V}_{L-2}\cdots\tilde{V}_2\tilde{V}_1,\\
V_{n}={}&U_{n+1}^\dagger U_{n}^{\vphantom{\dagger}},\;\;\tilde{V}_n={\cal P}V_n{\cal P}.
\end{align}
\end{subequations}
The two matrices $Q_L$ and $Q'_L$ have the same singular values. Furthermore, the $L-1$ matrices $V_1,V_2,\ldots V_{L-1}$ are independent and Haar distributed, so statistically equivalent to the set $U_1,U_2,\ldots U_{L-1}$. Hence $Q_L$ and $C_{L-1}$ have the same singular-value statistics.

\subsection{Gaussian approximation}

The advantage of the projector formulation is that one can easily implement the large-$N$ Gaussian approximation. Averages of products of the random vectors $u_i^{(n)}$ contain terms of different order in $N$. To leading order we may approximate the elements by independent Gaussian variables of zero mean and variance $1/N$. Moments can then be computed simply as sums of products of pairwise averages (Wick's theorem).

We take the large-$N$ limit at fixed truncation depth $\delta N$. For truncated unitary matrices this is the regime of \textit{weak non-unitarity} \cite{Ake14,Fyo99,Zyc00}. The product length $L$ may vary, we will in particular be interested in the double-scaling limit \cite{Ake15,Jia17}, where both $N$ and $L$ tend to infinity at fixed ratio $\tau=L\delta N/N$.

\section{Recursion relations}
\label{sec_recursion}

In a recursive approach we treat the number of projectors as a variable $n\in\mathbb{N}$, and consider the averages
\begin{equation}
S_p(n)=\frac{1}{N}\,\mathbb{E}\bigl[\operatorname{Tr} B_n^p],\;\;B_n\equiv Q_n^{\vphantom\dagger}Q_n^\dagger.
\end{equation}
The initial step of the recursion is $S_p(0)=1$. To increment $n$, we note that
\begin{align}
\operatorname{Tr}B_{n+1}^p={}&\operatorname{Tr}(P_{n+1} B_{n} P_{n+1})^p=\operatorname{Tr}(P_{n+1} B_{n})^p\nonumber\\
={}&\operatorname{Tr}\biggl(B_{n}-\sum_{i=1}^{\delta N}|u_i^{(n+1)}\rangle\langle u_i^{(n+1)}| B_{n}\biggr)^p
\end{align}

Since $B_n$ is independent of the vectors $v_i^{(n+1)}$, we can perform the Gaussian average over these vectors separately. We find
\begin{widetext}
\begin{align}
S_{p}(n+1)=S_p(n)+\frac{\delta N}{N}\underset{\sum_i x_i\neq 0}{\sum_{x_1=0}^1\cdots\sum_{x_p=0}^1}(-1)^{\sum_{i}x_i}\prod_{i=1}^p S_{d_i}(n)^{x_i}+{\cal O}(1/N^2),
\end{align}
where $d_i\in\{1,2,\ldots p\}$ is the smallest integer $\delta$ such that $x_{i+\delta\bmod(p)} =1$. 
For example, the first four recursion relations are
\begin{subequations}
\begin{align}
S_1(n+1)={}&(1-\delta N/N)S_1(n),\\
S_2(n+1) ={}& (1-2\delta N/N) S_2(n) + (\delta N/N)S_1(n)^2,\\
S_3(n+1) ={}& (1-3\delta N/N) S_3(n) +(\delta N/N)\left[- S_1(n)^3 + 3 S_1(n) S_2(n)\right],\\
S_4(n+1) ={}& (1-4\delta N/N) S_4(n) + (\delta N/N)\left[ S_1(n)^4 - 4 S_1(n)^2 S_2(n)+ 2S_2(n)^2 + 4S_1(n)S_3(n)\right].
\end{align}
\end{subequations}
\end{widetext}

These equations can be solved sequentially, the recursion relation for $S_p(n)$ is first-order linear once the recursion relations for $S_q(n)$ with $q<p$ have been solved.

\section{Double-scaling limit}
\label{sec_scaling}

\subsection{Singular-value moments of given order \textit{p}}

For $L\rightarrow\infty$ at fixed $N$ the moments decay exponentially, $S_p(L)\propto e^{pL\ln(1-\delta N/N)}$. The double-scaling limit $L,N\rightarrow\infty$ at fixed $\tau=L\delta N/N$ shows a more varied behavior. We find
\begin{equation}
\lim_{L=\tau N/\delta N\rightarrow\infty}S_p(L)=\frac{1}{\Gamma(p)}e^{-p\tau}G_p(\tau),\label{SGrelation}
\end{equation}
with $G_p(\tau)$ a polynomial in $\tau$ of degree $p-1$, given for $p\in\{1,2,3,4,5,6\}$ by
\begin{align}
&\bigl\{1, \tau+1,3 \tau^2+4 \tau+2, 16 \tau^3+24 \tau^2+18 \tau+6, \nonumber\\
&125 \tau^4+200 \tau^3+180 \tau^2+96 \tau+24,\label{Gptausmallp}\\
&1296 \tau^5+2160 \tau^4+2160 \tau^3+1440 \tau^2+600 \tau+120\bigr\}.\nonumber
\end{align}
This pattern matches the Erlang delay function \cite{Erlang},
\begin{subequations}
\label{GErlang}
\begin{align}
&G_p(\tau)=(1-\tau)(p-1)!\sum_{i=0}^{p-1} \frac{(p\tau)^i}{i!} + \tau^p p^{p-1}\label{GErlanga}\\
&\qquad=e^{p\tau}(1-\tau)(p-1)\Gamma(p-1, p\tau) + (p\tau)^{p-1},\label{GErlangb}
\end{align}
\end{subequations}
with $\Gamma(x,y)$ the incomplete Gamma function. We assume the correspondence between Eqs.\ \eqref{Gptausmallp} and \eqref{GErlang} holds for all $p$, although we have no insight why a delay function from queueing theory \cite{Coo81} would appear in the present context.

\begin{figure}[tb]
\centerline{\includegraphics[width=1\linewidth]{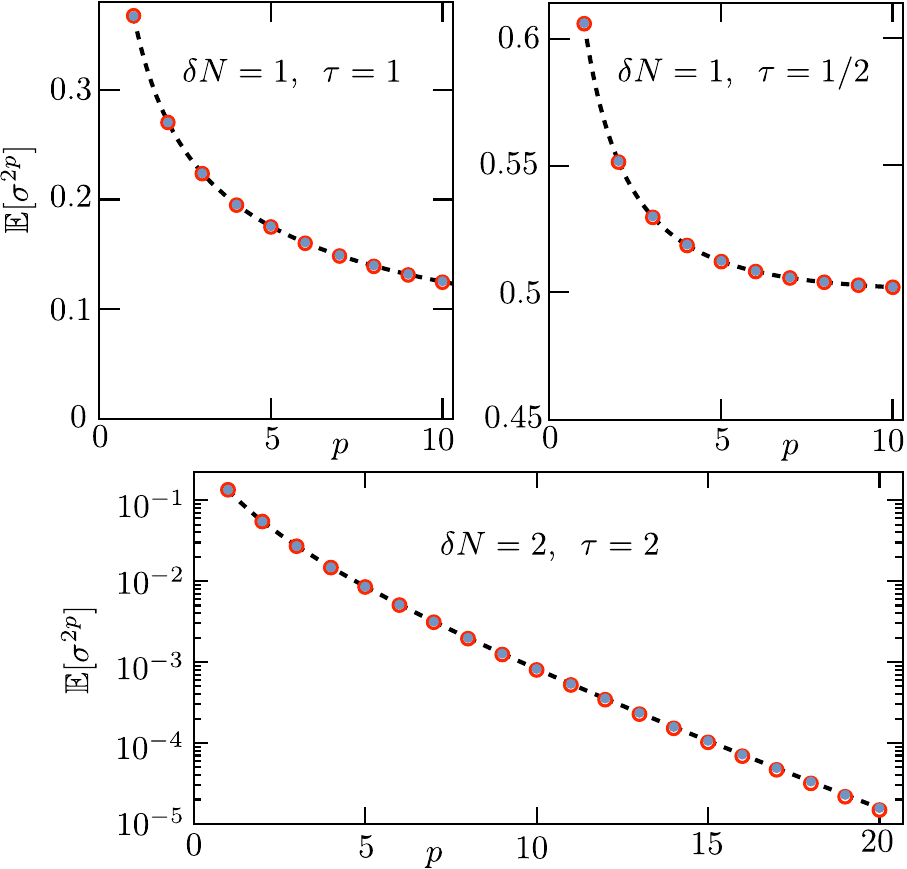}}
\caption{Data points: Singular-value moments, $N^{-1}\operatorname{Tr}(C_L^{\vphantom\dagger}C_L^\dagger)^p$, of the product of $L$ truncated random matrices from the unitary group $\mathrm{U}(N)$ (open red markers) or from the orthogonal group $\mathrm{O}(N)$ (closed blue markers). In each panel the matrix size $N=300$ and the data is averaged over 20 realizations. The three panels correspond to different truncation depths ($\delta N=1$ or $\delta N=2$ removed rows and columns) and to different lengths $L=\tau N/\delta N-1$ of the product \cite{note2}. Dashed curves: The formula \eqref{SGrelation} from the double-scaling limit.
}
\label{fig_moments}
\end{figure}

As a check, we show in Fig.\ \ref{fig_moments} that our analytical result \eqref{SGrelation} for the double-scaling limit agrees very well with a numerical average over random-matrix products. At the level of the  Gaussian approximation it does not matter whether the projection vectors are real or complex, Wick's theorem works in the same way for both, so we expect the large-$N$ singular-value statistics to be the same for complex unitary and real orthogonal matrix products. This is indeed confirmed by the numerics (compare blue and red data points).

\subsection{Large-\textit{p} asymptotics}

The large-$p$ asymptotics of the singular-value moments \eqref{SGrelation} can be obtained with the help of Eq.\ \eqref{GErlang} and the asymptotics of the incomplete Gamma function $\Gamma(x,y)$ when both arguments are large \cite{Fer05}. 

We can distinguish three regimes, depending on whether $\tau$ is smaller, equal, or larger than unity:
\begin{align}
&\lim_{L=\tau N/\delta N\rightarrow\infty}S_p(L)\underset{p\gg 1}{\sim}\nonumber\\
&\sim\begin{cases}
1-\tau&\text{if}\;\;0<\tau<1,\\
(2\pi p)^{-1/2}&\text{if}\;\;\tau=1,\\
p^{-3/2}e^{-(\tau-1-\ln\tau)p}&\text{if}\;\;\tau>1.
\end{cases}\label{momentsasymptotics}
\end{align}

This asymptotics can be interpreted in terms of the von-Neumann-Halperin theorem of alternating projections \cite{Neu49,Hal62,Gin18}: For large $L$ the projector product $Q_L$ converges to the projector onto the intersection of the $(N-\delta N)$-dimensional subspaces of the $L$ individual projections in $Q_L$. If $L\delta N<N$ this subspace can be expected to have rank $N-L\delta N=N(1-\tau)$, so a fraction $1-\tau$ of the singular values of $Q_L$ equals 1 and the rest equals 0. The moments then become $p$-independent, equal to $1-\tau$. If $\tau>1$ the intersection of the subspaces can be expected to be empty, with exponentially small singular values as a result. The case $\tau=1$ is the threshold between empty and non-empty subspace intersection, with a power law decay.

\section{Singular value density}
\label{sec_density}

\subsection{Reconstruction by analytical continuation}

The singular value density can be reconstructed from the moments $\langle\sigma^{2p}\rangle$ without going through the moment generating function, because with the help of Eq.\ \eqref{GErlangb} the result \eqref{SGrelation} for the moments $S_p(L)$ can be continued analytically to arbitrary complex $p$. 

If we change variables $\sigma^2=e^{-\lambda}$, with exponent $\lambda>0$, then the density ${\cal D}(\lambda)$ is related to $S_p(L)$ by Laplace transformation,
\begin{equation}
S_p(L)=\int_0^\infty e^{-p\lambda}{\cal D}(\lambda)\,d\lambda.
\end{equation}
The density can then be obtained by numerical inversion of the Laplace transform,
\begin{equation}
{\cal D}(\lambda)=\frac{1}{2\pi i}\int_{c-i\infty}^{c+i\infty}S_p(L)e^{p\lambda}\,dp.\label{Dlambda}
\end{equation}

\begin{figure}[tb]
\centerline{\includegraphics[width=0.8\linewidth]{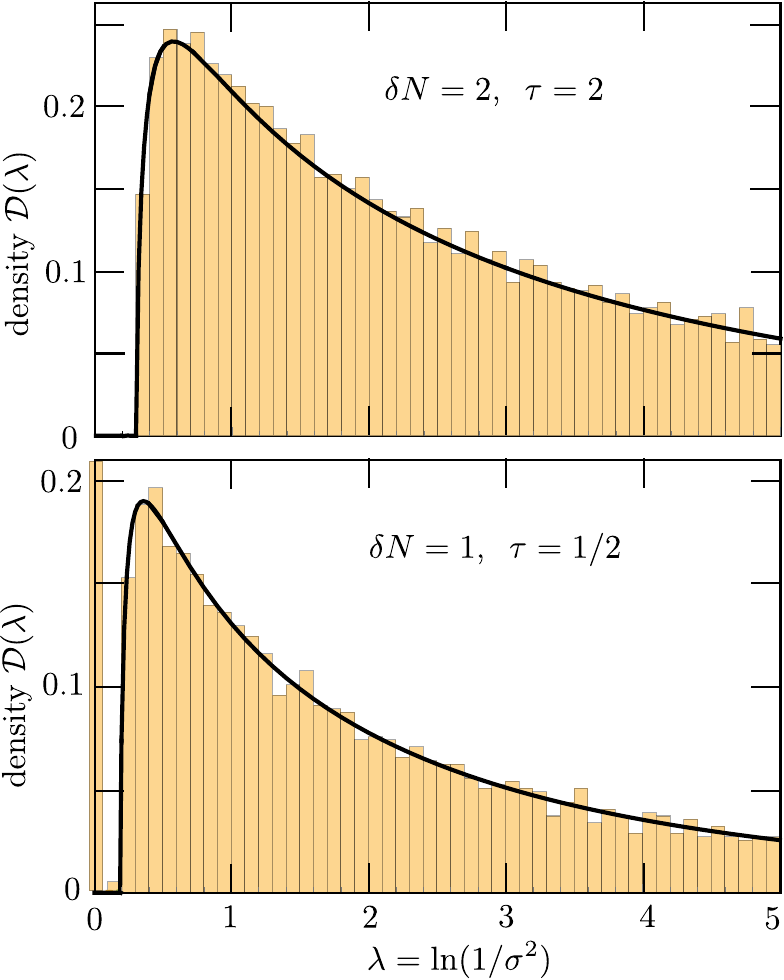}}
\caption{Histogram: Singular-value density ${\cal D}$ (in terms of the variable $\lambda=-\ln\sigma^2)$ for the product of $L$ truncated random matrices from the unitary group $\mathrm{U}(N)$ (with $N=300$), averaged over $20$ realizations. Black curve: ${\cal D}(\lambda)$ calculated from the inverse Laplace transform \eqref{Dlambda}. The two panels correspond to two different values of the truncation depth $\delta N$  and product length $L=\tau N/\delta N-1$. For $\tau<1$ the density has a delta-function peak at $\lambda=0$. The positive $\lambda$'s have a minimal value $\lambda_{\rm min}=\tau-1-\ln\tau$.
}
\label{fig_density}
\end{figure}

Results are shown in Fig.\ \ref{fig_density}. The $p^{-3/2}e^{-p\lambda_{\rm min}}$ large-$p$ tail \eqref{momentsasymptotics} produces the square-root lower limit for $\tau> 1$,
\begin{equation}
{\cal D}(\lambda)\propto \sqrt{\lambda-\lambda_{\rm min}},\;\;\lambda_{\rm min}=\tau-1-\ln\tau,
\end{equation}
which for $\tau<1$ is supplemented by a delta function $(1-\tau)\delta(\lambda)$ at unit singular value.

\begin{figure}[tb]
\centerline{\includegraphics[width=0.8\linewidth]{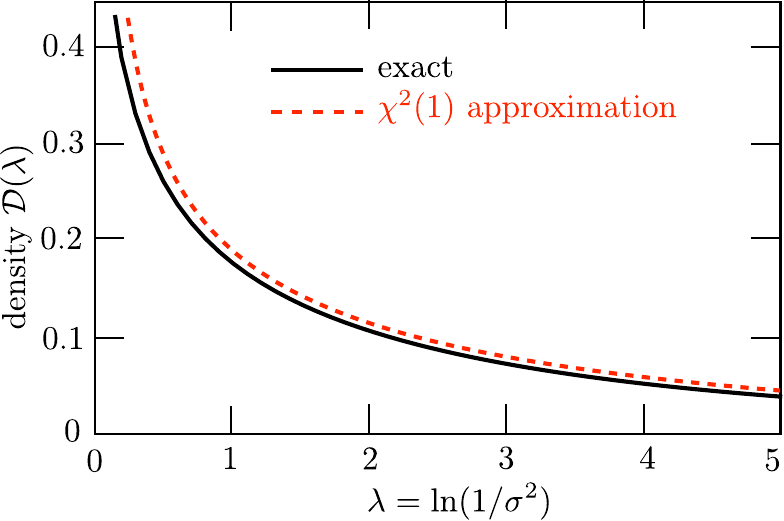}}
\caption{Comparison for $\tau=1$ of the exact probability density function, as given by the inverse Laplace transform \eqref{Dlambda}, and the approximation \eqref{sigma2PDF} by a chi-square distribution.
}
\label{fig_PDF}
\end{figure}

The threshold case $\tau=1$ is qualitatively different, as can be seen directly from the moments \eqref{SGrelation}. We may approximate
\begin{equation}
\langle\sigma^{2p}\rangle=(p^p/p!)e^{-p}\approx(1+2\pi p)^{-1/2}.\label{tauis1moments}
\end{equation}
The error in this approximation is less than $1\%$ for all $p\in\mathbb{N}$. The set of moments $(1+2\pi p)^{-1/2}$ characterizes the exponentiated chi-square distribution with one degree of freedom,
\begin{equation}
\sigma^2=e^{-\pi Y},\;\;Y\sim\chi^2(1).\label{sigma2PDF}
\end{equation}
As shown in Fig.\ \ref{fig_PDF}, the $\chi^2(1)$ probability density function describes quite well the singular value density for $\tau=1$.

\section{Entropy}
\label{sec_entropy}

The Hermitian trace-one matrix
\begin{equation}
\rho_L=\frac{C_L^{\vphantom\dagger}C_L^\dagger}{\operatorname{Tr}C_L^{\vphantom\dagger}C_L^\dagger}\label{rhoLdef}
\end{equation}
can be interpreted as the density matrix of a quantum system \cite{Bul24}. In that context the von Neumann entropy ${\cal S}_L=-\operatorname{Tr}\,\rho_L\ln\rho_L$ is a measure of the purity of the state: ${\cal S}_L=0$ for a pure state and ${\cal S}_L=\ln N$ for a maximally mixed state. Let us calculate this quantity.

It is helpful to start with the more general R\'{e}nyi entropy
\begin{equation}
{\cal E}_\alpha(L)=\frac{1}{1-\alpha}\ln\operatorname{Tr}\rho_L^\alpha,
\end{equation}
defined for $\alpha>0$, with the von Neumann entropy as the $\alpha\rightarrow 1$ limit,
\begin{equation}
\lim_{\alpha\rightarrow 1}{\cal E}_\alpha(L)=-\operatorname{Tr}\,\rho_L\ln\rho_L\equiv {\cal S}_L.
\end{equation}
Substitution of Eq.\ \eqref{rhoLdef} gives
\begin{equation}
{\cal E}_\alpha(L)=\frac{1}{1-\alpha}\ln\left[\frac{\operatorname{Tr}(C_L^{\vphantom\dagger}C_L^\dagger)^\alpha}{\bigl(\operatorname{Tr}C_L^{\vphantom\dagger}C_L^\dagger\bigr)^\alpha}\right]
\end{equation}
For large $N$ we may approximate each trace by its expectation value, $\operatorname{Tr}(C_L^{\vphantom\dagger}C_L^\dagger)^p\approx NS_p(L)$, thus expressing the R\'{e}nyi entropy in terms of singular-value moments,
\begin{equation}
{\cal E}_\alpha(L)\approx\ln N+\frac{1}{1-\alpha}\ln\left[\frac{S_{\alpha}(L)}{S_1(L)^\alpha}\right].
\end{equation}

The difference 
\begin{equation}
\delta{\cal E}_\alpha(L)={\cal E}_\alpha(L)-\ln N
\end{equation}
quantifies how far the density matrix is from being maximally mixed. The double-scaling limit \eqref{SGrelation} gives
\begin{equation}
\lim_{L=\tau N/\delta N\rightarrow\infty}\delta{\cal E}_\alpha(L)=\frac{1}{1-\alpha}\ln\left[ \frac{G_\alpha(\tau)}{\Gamma(\alpha)}\right].
\end{equation}

\begin{figure}[tb]
\centerline{\includegraphics[width=0.8\linewidth]{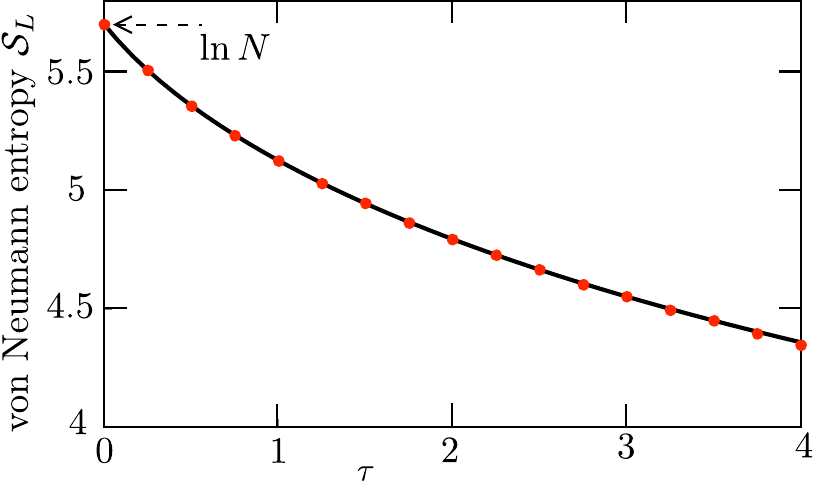}}
\caption{Data points: von Neumann entropy ${\cal S}_L$ of the product of $L$ truncated random matrices from the unitary group $\mathrm{U}(N)$, with $N=300$, truncation depth $\delta N=1$, and different $L=\tau N/\delta N-1$. This is data for a single realization (no averaging). The solid curve is the result \eqref{entropylimit} in the double-scaling limit.
}
\label{fig_entropy}
\end{figure}

To obtain the von Neumann entropy we use Eq.\ \eqref{GErlangb} for the analytic continuation of the function $G_p(\tau)$ to non-integer $p$. A series expansion around $p=\alpha=1$ then gives the result
\begin{align}
&\lim_{L=\tau N/\delta N\rightarrow\infty}({\cal S}_L-\ln N)=\lim_{\alpha\rightarrow 1}\frac{1}{1-\alpha}\ln\left[ \frac{G_\alpha(\tau)}{\Gamma(\alpha)}\right]\nonumber\\
&\qquad=-\ln \tau+e^{\tau} (\tau-1) \Gamma (0,\tau)-\gamma_{\rm Euler}.\label{entropylimit}
\end{align}
Fig.\ \ref{fig_entropy} shows that Eq.\ \eqref{entropylimit} is very close to the numerical result for large $N$.

While the precise functional form \eqref{entropylimit} is model dependent, the asymptotics for small and large $\tau$,
\begin{equation}
\lim_{L=\tau N/\delta N\rightarrow\infty}({\cal S}_L-\ln N)=\begin{cases}
-\tau&\text{if}\;\;\tau\ll 1,\\
1-\ln \tau-\gamma_{\rm Euler} &\text{if}\;\;\tau\gg 1,
\end{cases}
\end{equation}
agrees with Refs.\ \onlinecite{Bul24,Ger24,DeL24}, where different (non-projective) weak-measurement models were studied \cite{note3}. We note that our large-$\tau$ regime is limited by $\tau\ll N$, the regime $\tau\gtrsim N$ studied in Refs.\ \onlinecite{Bul24,Ger24,DeL24} is not accessible in the double-scaling limit.

\section{Conclusion}
\label{sec_conclude}

The central technical result of our study is the expression \eqref{SGrelation} for singular value moments $\langle\sigma^{2p}\rangle$ of a product of truncated unitary matrices, in the double-scaling limit that both the matrix dimension $N$ and the number of independent matrices $L$ goes to infinity, at fixed ratio $\tau=L\delta N/N$ (with $\delta N\ll N$ the truncation depth). Because this expression can be analytically continued to complex powers $p$, it directly provides the full eigenvalue density function by Laplace transform inversion. 

We have applied this result to a calculation of the von Neumann entropy ${\cal S}$ of the density matrix corresponding to the random contraction process, as a simple model for measurement-induced purification in a monitored quantum circuit \cite{Li18,Cha19,Ski19,Szy19,Cho20,Gul20}. Our findings are consistent with recent work \cite{Bul24,Ger24,DeL24} on this problem, although our projective implementation of weak measurements is different, confirming the universality of the logarithmic decay ${\cal S}\approx\ln (N/\tau)$ for $1\ll\tau\ll N$.

From a more general perspective this is a contribution to the random-matrix theory of products of truncated unitary (or orthogonal) matrices, with an extensive literature \cite{Ake14,Ake15,Ibs14,Adh16,Kie16,For18,Adh20} and a variety of applications in statistical physics and data analysis. 

One application that we mention is to projection algorithms for image reconstruction \cite{Fer24,Pop18,Kan21} (Kaczmarz-type algorithms). One then studies a product of $N$ rank-$(N-1)$ projectors,
\begin{equation}
Q_N=\prod_{n=1}^N(\openone-|u^{(n)}\rangle\langle u^{(n)}|),
\end{equation}
with $u^{(1)},u^{(2)},\ldots u^{(N)}$ a set of real unit vectors of length $N$. If we assume that the unit vectors are independently and isotropically distributed, this corresponds to our $\tau=1$ orthogonal matrix product, with singular-value moments given by Eq.\ \eqref{tauis1moments},
\begin{equation}
\lim_{N\rightarrow\infty}\frac{1}{N}\mathbb{E}\bigl[\operatorname{Tr} (Q_N^{\vphantom{\top}}Q_N^\top)^p\bigr]=\frac{(p/e)^p}{p!}.
\end{equation}
As shown in Fig. \ref{fig_PDF}, this is close to an exponentiated chi-square distribution. Is there a direct way to see this correspondence?

\acknowledgments

This project was motivated by \href{https://mathoverflow.net/q/475439/11260}{a question} at MathOverflow from Yaroslav Bulatov, and benefited from the insights of Xiangxiang Xu.


\begin{thebibliography}{99}
\bibitem{Wil18} M. Wilkinson and J. Grant, \textit{A matrix contraction process}, J. Phys. A \textbf{51}, 105002 (2018).
\bibitem{Neu49} J. von Neumann, \textit{On rings of operators. Reduction theory}, Ann. Math. \textbf{50}, 401 (1949).
\bibitem{Hal62} I. Halperin, \textit{The product of projection operators}, Acta Sci. Math. (Szeged) \textbf{23}, 96 (1962).
\bibitem{Gin18} O. Ginat, \textit{The method of alternating projections}, arXiv:1809.05858.
\bibitem{Cri93} A. Crisanti, G. Paladin, and A. Vulpiani, \textit{Products of Random Matrices in Statistical Physics} (Springer, 1993).
\bibitem{Ake14} G. Akemann, Z. Burda, M. Kieburg, and T. Nagao, \textit{Universal microscopic correlation functions for products of truncated unitary matrices}, J. Phys. A \textbf{47}, 255202 (2014).
\bibitem{Ake15} G. Akemann and J. R. Ipsen, \textit{Recent exact and asymptotic results for products of independent random matrices}, Acta Phys. Polonica B \textbf{46}, 1747 (2015).
\bibitem{Ibs14} J. R. Ipsen and M. Kieburg, \textit{Weak commutation relations and eigenvalue statistics for products of rectangular random matrices}, Phys. Rev. E \textbf{89}, 032106 (2014).
\bibitem{Adh16} K. Adhikari, N. Kishore Reddy, T. Ram Reddy, and K. Saha, \textit{Determinantal point processes in the plane from products of random matrices}, Ann. Inst. H. Poincar\'{e} Probab. Statist. \textbf{52}, 16 (2016).
\bibitem{Kie16} M. Kieburg, A. B. J. Kuijlaars, and D. Stivigny, \textit{Singular value statistics of matrix products with truncated unitary matrices}, Int. Math. Res. Notices \textbf{2016}, 3392 (2016).
\bibitem{For18} P. J. Forrester and S. Kumar, \textit{The probability that all eigenvalues are real for products of truncated real orthogonal random matrices}, J. Theor. Prob. \textbf{31}, 2056 (2018).
\bibitem{Adh20} K. Adhikari and A. Bose, \textit{Limiting spectral distribution of the product of truncated Haar unitary matrices}, 
Random Matrices: Theory and Appl. \textbf{09}, 2050002 (2020).
\bibitem{Gia23} G. Giacomin, \href{https://www.math.unipd.it/~giacomin/notesRM-SM_24_10_2024.pdf}{Random Matrix Products and the Statistical Mechanics of Disordered Systems} (2024).
\bibitem{note1} To truncate the matrix, we might remove the first $\delta N$ rows and columns, but it is convenient to keep the size of $U_n$ and $\tilde{U}_n$ the same, so instead we set these rows and columns to zero.
\bibitem{Bul24} V. B. Bulchandani, S. L. Sondhi, and J. T. Chalker, \textit{Random-matrix models of monitored quantum circuits}, J. Stat. Phys. \textbf{191}, 55 (2024)
\bibitem{Ger24} F. Gerbino, P. Le Doussal, G. Giachetti, and A. De Luca, \textit{A Dyson brownian motion model for weak measurements in chaotic quantum systems}, Quantum Rep. \textbf{6}, 200 (2024).
\bibitem{DeL24} A. De Luca, C. Liu, A. Nahum, and T. Zhou, \textit{Universality classes for purification in nonunitary quantum processes}, arXiv:2312.17744.
\bibitem{Li18} Yaodong Li, Xiao Chen, and Matthew P. A. Fisher, \textit{Quantum Zeno effect and the many-body entanglement transition}, Phys. Rev. B \textbf{98}, 205136 (2018).
\bibitem{Cha19} A. Chan, R. M. Nandkishore, M. Pretko, and G. Smith, Phys. Rev. B \textbf{99}, 224307 (2019).
\bibitem{Ski19} B. Skinner, J. Ruhman, and A. Nahum, \textit{Measurement-induced phase transitions in the dynamics of entanglement}, Phys. Rev. X \textbf{9}, 031009 (2019).
\bibitem{Szy19} M. Szyniszewski, A. Romito, and H. Schomerus, \textit{Entanglement transition from variable-strength weak measurements}, Phys. Rev. B \textbf{100}, 064204 (2019).
\bibitem{Cho20} Soonwon Choi, Yimu Bao, Xiao-Liang Qi, and Ehud Altman, \textit{Quantum error correction in scrambling dynamics and measurement-induced phase transition}, Phys. Rev. Lett. \textbf{125}, 030505 (2020).
\bibitem{Gul20} M. J. Gullans and D. A. Huse, \textit{Dynamical purification phase transition induced by quantum measurements}, Phys. Rev. X \textbf{10} 041020 (2020).
\bibitem{Coo81} R. B. Cooper, \textit{Introduction to Queueing Theory} (North Holland, 1981).
\bibitem{Fyo99} Y. V. Fyodorov and B. A. Khoruzhenko, \textit{Systematic analytical approach to correlation functions of resonances in quantum chaotic scattering}, Phys. Rev. Lett. \textbf{83}, 65 (1999).
\bibitem{Zyc00} K. Zyczkowski and H.-J. Sommers, \textit{Truncations of random unitary matrices}, J. Phys. A \textbf{33}, 2045 (2000).
\bibitem{Jia17} Tiefeng Jiang and Yongcheng Qi, \textit{Spectral radii of large non-Hermitian random matrices}, J. Theor. Prob. \textbf{30}, 326 (2017).
\bibitem{Erlang} The Erlang delay formula is $\propto 1/G_p(\tau)$, see series A122525 at the \href{https://oeis.org/A122525}{OEIS database}.
\bibitem{Coo81} R. B. Cooper, \textit{Introduction to Queueing Theory} (Elsevier, 1981).
\bibitem{note2} Because of the correspondence of $C_L$ with $Q_{L+1}$ we compute the parameter $\tau$ as $\tau=(L+1)\delta N/N$ when comparing the numerics for random-matrix products with the analytics for random-projector products. The $+1$ slightly improves the agreement.
\bibitem{Fer05} C. Ferreira, J. L\'{o}pez, and E. P\'{e}rez Sinus\'{\i}a, \textit{Incomplete gamma functions for large values of their variables}, Adv. Appl. Math. \textbf{34}, 467 (2005).
\bibitem{note3} Our projective implementation of a weak measurement corresponds to a collective measurement of the whole system. For example, if the Hilbert space of dimension $N=2^{\cal N}$ is constructed from ${\cal N}$ spin-$1/2$ degrees of freedom, a projection onto a $N-1$-dimensional subspace is realized by a threshold measurement that checks if the total spin is less than its maximal value of ${\cal N}/2$ --- without measuring the individual spins.
\bibitem{Fer24} I. A. Ferreira, J. A. Acebr\'{o}n, and J. Monteiro, \textit{Survey of a class of iterative row-action methods: The Kaczmarz method}, arXiv:2401.02842.
\bibitem{Pop18} C. Popa, \textit{Convergence rates for Kaczmarz-type algorithms}, Numer. Algor. \textbf{79}, 1 (2018).
\bibitem{Kan21} Chuan-gang Kang, \textit{Convergence rates of the Kaczmarz-Tanabe method for linear systems}, J. Comp. Appl. Math. \textbf{394}, 113577 (2021).

\end{thebibliography}
\end{document}